\documentclass{article}
\usepackage{spconf,amsmath,graphicx,booktabs}
\usepackage[usenames,dvipsnames]{xcolor}

\title{Latent space representation for multi-target speaker detection and identification with a sparse dataset using Triplet neural networks}
\name{Kin Wai Cheuk$^{1,2}$, Balamurali B T$^1$, Gemma Roig$^1$, Dorien Herremans$^{1,2}$}
\address{$^1$Singapore University of Technology and Design\\
$^2$Institute of High Performance Computing, A*STAR}

\usepackage[numbers]{natbib}

\usepackage{fancyhdr}

\fancypagestyle{plain}{%
	\fancyhf{}%
	\fancyfoot[L]{\footnotesize \center \emph{Preprint accepted for publication in the Proceedings of IEEE Automatic Speech Recognition and Understanding Workshop (ASRU 2019). Singapore. 2019. \\ -\thepage-}}%
}
\begin{document}
%
\pagestyle{fancy}

\maketitle
\begin{abstract}
We present an approach to tackle the speaker recognition problem using Triplet Neural Networks. Currently, the $i$-vector representation with probabilistic linear discriminant analysis (PLDA) is the most commonly used technique to solve this problem, due to high classification accuracy with a relatively short computation time. In this paper, we explore a neural network approach, namely Triplet Neural Networks (TNNs), to built a latent space for different classifiers to solve the Multi-Target Speaker Detection and Identification Challenge Evaluation 2018 (MCE 2018) dataset. This training set contains $i$-vectors from 3,631 speakers, with only 3 samples for each speaker, thus making speaker recognition a challenging task.  When using the train and development set for training both the TNN and baseline model (i.e., similarity evaluation directly on the $i$-vector representation), our proposed model outperforms the baseline
by 23\%. When reducing the training data to only using the train set, our method results in 309 confusions for the Multi-target speaker identification task, which is 46\% better than the baseline model. These results show that the representational power of TNNs is especially evident when training on small datasets with few instances available per class. 
\end{abstract}
\section{Introduction}
\label{sec:intro}

This paper explores the performance of Triplet Neural Networks (TNNs) on the speaker detection and identification task in a setting with a small number of training samples per class (3 samples for each speaker). TNNs were first proposed by Google in 2015~\cite{schroff2015facenet} and have shown to be particularly efficient for solving classification problems that only have a few training samples available per class~\cite{ye2018deep}. Prior research has reported the effectiveness of TNNs for this type of dataset in the area of computer vision~\cite{ye2018deep, chen2016deep,cheng2016person,su2016deep,wang2016joint,ding2015deep}. Recently, this technique has been adapted to the audio research domain, for tackling tasks such as speaker diarization~\cite{song2018triplet, le2017triplet}, speaker verification~\cite{zhang2018text, li2018deep}, speaker identifications~\cite{li2017deep}, and speaker change detection~\cite{sari2019pre}. In this paper, our main focus is on speaker identification, which we test using  the dataset from the Multi-Target Speaker Detection and Identification Challenge~\cite{cheuk2018mce}, a competition focused on the task of automatic blacklisted speaker detection using audio recordings. This dataset\footnote{http://MCE 2018.org/} consists of $i$-vectors~\cite{shon2018mce} of length 600. Each of these $i$-vectors corresponds to a real-world telephone conversation by customers and agents from a call-center.  Much of the existing research on multi-target speaker detection uses larger datasets than the aforementioned to train their models, for example in~\cite{li2018deep,zhang2018text}, they have 300 utterances for each speaker; other work by \citet{li2017deep} uses 10-50 utterances for each speaker. The MCE 2018 dataset~\cite{shon2018mce}, however, is challenging as there are only three samples for each speaker. 

There is a plethora of algorithms for solving the multi-target speaker detection problem, some of the most well known methods include Gaussian Mixture Model with Universal non-blacklisted Model (GMM-UBM)~\cite{hasan2011study}, Joint Factor Analysis (JFA)~\cite{kenny2005joint}, and $i$-vectors~\cite{soong1987report,garcia2011analysis} with linear probabilistic linear discriminant analysis (LDA)~\cite{izenman2013linear,ioffe2006probabilistic,burget2011discriminatively}. Despite the successful results achieved by the above mentioned methods, neural network approaches are becoming ever more popular.  We explore the power of TNNs on classification with a small dataset, i.e., MCE 2018~\cite{shon2018mce}.

In this paper, we tackle two tasks, using the same TNN network architecture. Task~1 is a speaker verification task that consists of classifying $i$-vectors between blacklisted and non-blacklisted speakers (i.e., to predict whether a given unknown recording is spoken by a blacklisted speaker or not). Task~2 is a speaker identification task whereby each of the 3,631 unique blacklisted speakers form a separate class (i.e., to identify which specific blacklisted speaker was talking). The MCE 2018~\cite{shon2018mce} dataset does not provide the original audio files, but only the calculated $i$-vectors. We will therefore use these as input to the TNN model. Since only the $i$-vectors are made available, there is no way to obtain other representations such as $d$-vector and $x$-vector from the same MCE 2018 dataset at the moment.  We propose a hybrid method that integrates the learned TNN representation with a classification algorithm, to solve both Task 1 and Task 2 of the MCE 2018 competition~\cite{hoffer2015deep,schroff2015facenet, cheukblacklisted}. The advantage of our TNN approach over existing algorithms is that TNNs are known to be especially efficient for classification tasks with only a few training samples per class~\cite{ye2018deep}. Given that the MCE dataset contains only three samples per class for Task~2, we expect our proposed TNN based models to reach high accuracy.

\section{System Description}
\label{sec:Description}

\begin{figure}[h]
    \centering
    \includegraphics[width=0.49\textwidth]{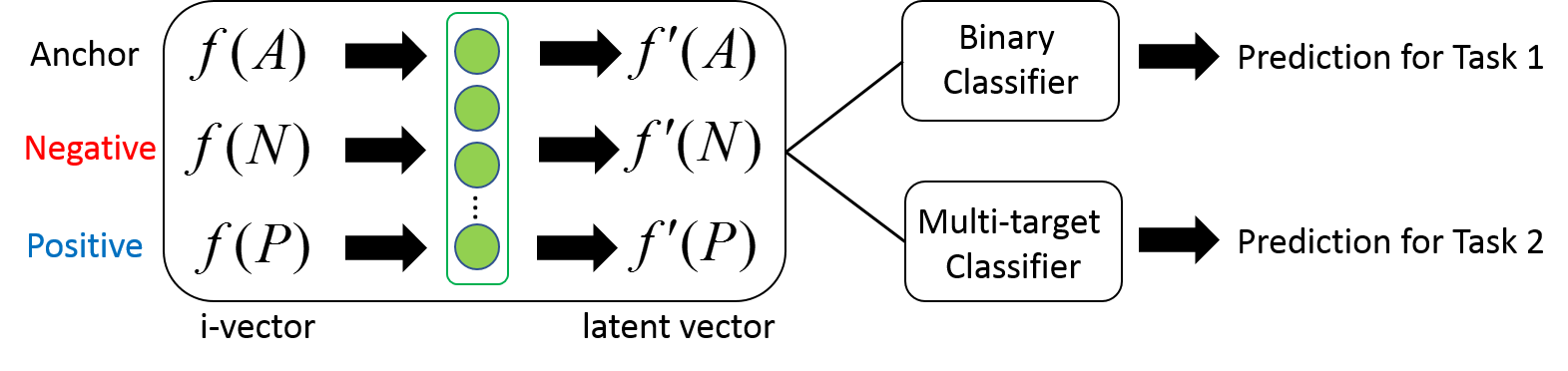}
    \caption{Our proposed hybrid classification system, which integrates a TNN and a classifier. }
    \label{fig:training}
\end{figure}

In order to tackle the two aforementioned classification tasks, we propose to use a Triplet Neural Network (TNN) that is trained to maximize the distance between elements of different classes, and minimize the distance between elements of the same class. In our case, the classes are non-blacklisted and blacklisted speakers (for Task~1) and individual blacklisted speakers (for Task~2). The TNN allows us to transform the original $i$-vector into a new latent space representation. This new representation can then be used to train a classifier for each of the two tasks respectively (see Figure~\ref{fig:training}). 


\subsection{Triplet Neural Networks}

The TNN was trained using input data in triplet form, A-P-N, whereby $A$ is the anchor $i$-vector, $P$ is a positive example $i$-vector which is of the same class as $A$, and $N$ is a negative example $i$-vector which is of a different class (non-blacklisted and blacklisted speakers for Task~1, individual blacklisted speakers for Task~2) than $A$. The loss function used for training the TNN is shown in Equation~\ref{eq:distance}, and aims at maximizing the distance between instances of different classes when represented in a newly learned multi-dimensional space, while minimizing the distance between instances of the same class. It is important to note that the TNN uses the same network (with exactly the same weights) for each of the three inputs. The training objective is to minimize the following function:   
\begin{equation}
  L(A,P,N) = \max(D(A,P) - D(A,N) + \alpha, 0),
  \label{eq:distance}
\end{equation}
whereby $L$ is the loss and $\alpha$ is the margin between clusters. In order to minimize $L$,  the Euclidean distance between the anchor and positive examples, $D(A,P)$ should be minimized, and the distance between the anchor and negative examples $D(A,N)$ maximized~\cite{hoffer2015deep}. The $\max$ operation prevents the loss from becoming negative. If the $D(A,P) - D(A,N)$ term should become negative, the $max$ operation will return 0, and the network will stop training.

Two different strategies for training the TNN were explored, as described below.

\begin{description}
\item [Method 1 (M1):]  Train one  TNN model for Task~1 and another one for Task~2. Then, use the corresponding trained model to transform the input data into the newly learned representation for Task 1 and Task 2, respectively.
\item [Method 2 (M2):]  Train the TNN weights for Task~2 only and apply the same TNN to transform the input data for both tasks into a newly learned representation.
\end{description}

Different architectures with varying numbers of neurons for the TNN were explored. A higher number of neurons (1,024 or 2,048) allows for a faster training convergence and better latent space representation, yet the training complexity for the classifiers also increases accordingly. A lower number of neurons causes information loss, resulting in lower classification accuracy. We also tested architectures with a varying number of layers. This showed that deeper models have a worse prediction performance, most likely due to our limited dataset. Based on our findings, a single fully connected layer TNN with 600 neurons strikes the balance between accuracy and model complexity, and was therefore chosen as our network architecture. ReLU was used as the activation function and Adam with learning rate $10^{-5}$  as the optimizer. A batch size of 1,024 triplets was selected and the network was trained for 100 epochs.


\subsection{Classification algorithms}
\label{sec:class}

Although TNNs learn a latent representation that better separates classes, they are not discriminative models. We still need to feed this representation into a classifier to predict the final class of an instance. Three classifiers were considered in this work, including a one-versus-all support vector machine (SVM) with Radial Basis Function (RBF) kernel~\cite{hearst1998support, steinwart2008support}, K-nearest neighbours (KNN)~\cite{keller1985fuzzy,weinberger2006distance}, and a cosine similarity classifier. Due to the large number of classes (3,631), it was not practically feasible to use grid search to find the optimal parameters for SVM. Instead, we set $C=1$ and $\gamma = 1/600$. For KNN, the number of neighbours was set to three (i.e., $k=3$), with standard Euclidean distance as the distance metric. The Python library scikit-learn was used to implement both of these classifiers~\cite{scikit-learn}. 

Finally, the cosine similarity classifier is identical to the one used by the MCE 2018 competition to calculate the baseline score. This classifier works by calculating the average of the $i$-vectors for each blacklisted speaker as~$x_{avg}^{(j)}$, whereby $j \in [0,3630]$ represents each of the different blacklisted speakers. The similarity score is then calculated by taking the dot product between the unknown $i$-vector, $X$, and the average $i$-vector per speaker. For every $i$-vector with unknown label, the cosine similarity is calculated calculated 3,631 times (i.e., for each possible blacklisted speaker). If the highest similarity score out of all the calculated scores for a given speaker is greater than a given threshold, we label the unknown $i$-vector as a blacklisted speaker. The corresponding speaker is chosen the be the one with the greatest similarity score. 

\section{Experimental set-up}

To evaluate the efficiency of our TNN approach on both speaker identification and blacklisted speaker detection, we performed a series of experiments on the MCE 2018 dataset. Information on the dataset and an in-depth understanding of both classification tasks are given below. 

\subsection{Dataset}

The MCE 2018 challenge provides a dataset that consists of recordings of conversations of both blacklisted and non-blacklisted speakers. These recordings are obtained from a call center, whereby some of the customers have been marked as blacklisted \cite{shon2018mce}. The exact reasons for blacklisting customers are not given, only the label (blacklisted and non-blacklisted) is provided for each recording. This dataset consists of a training set, development set, and test set, as shown in Table~\ref{tab:dataset}. It is important to note that the recordings of blacklisted conversations in the test set belong to the same speakers that are also present in the training and development set. The non-blacklisted conversations in the test set, however, belong to unknown speakers, thereby making evaluation on this particular set more challenging.

\begin{table}[h]
\small
\centering
\caption{Summary of the MCE 2018 dataset.  }
\label{tab:dataset}
\begin{tabular}{l|cc}
\toprule
Set& Unique bl id (\# samples)   &  bg id (\# samples) \\
\midrule
Training       & 3,631 (10,893) & 5,000 (30,952)\\ 
Development  & 3,631 (3,631)& 5,000 (5,000)\\
Test  & 3,631 (3,631)& Unknown (12,386)\\
\bottomrule
\end{tabular}
\footnotesize \flushleft
bl stands for blacklisted speaker, and bg for non-blacklisted speaker. The test set contains new non-blacklisted speakers that do not appear in the training and development set.
\end{table}

Two different experiments were undertaken as part of this investigation. In the first experiment, our proposed model was trained using the given training data, and evaluated using the development data (denoted as Set A). This set reflects the initial baseline provided by the MCE challenge before the test set was made available. In the second experiment, the model was trained using data from both the training and development set and was evaluated on the given test data (denoted as Set B).  These details are summarized  in Table~\ref{tab:datasplit}. We introduce the two tackled classification tasks in Section \ref{task1} and section \ref{task2} below.

\begin{table}[h]
\centering
\small
\caption{Dataset split. }
\label{tab:datasplit}
\begin{tabular}{l|cc}
\toprule
Set used for: & Set A    &  Set B  \\
\midrule
Training       & Training set & Training + Development set\\ 
Evaluation  & Development set& Test set\\
\bottomrule
\end{tabular}
\end{table}

\subsection{Task 1: Blacklisted speaker detection}\label{task1}

The aim of this task is to identify whether a given $i$-vector belongs to a blacklisted speaker or not. This is a binary classification problem for which we are predicting the label of a given recording. This label can either be 1 (blacklisted) or 0 (not-blacklised).

Given that the dataset consists of 3,631 unique blacklisted speakers (3 samples for each speaker) and 5,000 non-blacklisted speakers (at least 4 samples each), the number of possible A-P-N triplets that can be generated is huge. We randomly sampled from these possible combinations and trained the network for multiple epochs. 
For this task, the number of sampled triplets for each batch is 96,000. Using these samples, the network was trained for 30 epochs. After these 30 epochs, another set of 96,000 A-P-N triplets was sampled and training continued for 30 epochs. This process was repeated four times, such that the network converges and the loss function approaches zero. 
 
The final weights of the model were used to transform the original $i$-vectors into a new latent space. The different classifiers described in Section~\ref{sec:class} were then used to predict the likelihood of an $i$-vector belonging to a blacklisted speaker or not. The result is reported in terms of equal error rate (EER), a metric that finds a threshold whereby the false acceptance rate is equal to the false rejection rate. A lower EER indicates a higher classification accuracy.

\subsection{Task 2: Speaker identification}\label{task2}

Assuming that an unknown $i$-vector belongs to a blacklisted speaker, the aim of Task 2 is to find out the identity of the speaker from a dataset of 3,631 blacklisted speakers. This is a non-trivial task, as there are only three samples provided per speaker. 

For each batch, we sampled 1,000,000 triplet combinations which were used to train the TNN. After 5 epochs, a new batch was sampled to continue the TNN training (similar to the procedure for Task~1). This was repeated until the loss was near to zero. Task~2 was then evaluated using three classifiers (KNN, SVM, and cosine similarity).

In addition, we discovered that the TNN weights learned for this task can also be applied to Task 1 (Method 2, set B), resulting in a better prediction accuracy as discussed below. The evaluation of this task is done in terms of the number of confusions, i.e., misclassifications.

 
\section{Results and Discussion}

The performance of the proposed system is evaluated below for both Task 1 and 2.

\subsection{Task 1: Blacklisted speaker detection}

Tables~\ref{tab:task1} and~\ref{tab:task2} present a summary of the models' performance when evaluating on evaluation set A and B respectively (see Table~\ref{tab:datasplit}). On set A, our proposed TNN model, trained using Method 1 (i.e., by training the TNN of each task separately), combined with KNN outperforms all the other models with an EER as low as 0.84\%.

\begin{table}[ht]
\centering
\caption{Model performance on evaluation set A.  The TNN was trained using Method 1 (M1) and Method 2 (M2).}
\label{tab:task1}
\small
\begin{tabular}{l|ccc}
\toprule
& Task 1 (M1) & Task 1 (M2)   & Task 2 \\
& (EER) & (ERR) & (Confusions) \\ \midrule
TNN-cosine    & 1.22\% & 1.84\% & 380 \\ 
TNN-KNN & \pmb{0.84\%}& 2.34\% & 428 \\
TNN-SVM & 1.44\% &1.45\% & \pmb{358} \\
baseline & 2.00\%& N.A. & 444\\\bottomrule
\end{tabular}

\end{table}

When evaluating our models on evaluation set B, which contains unknown non-blacklisted speakers, the TNN-based methods are not able to outperform the MCE 2018 competition baseline. The best method using TNNs, is the TNN combined with cosine similarity classifier, which obtains an EER of 8.49\% when trained using Method 1. 

\begin{table}[ht]
\centering
\caption{Model performance on evaluation set B. The TNN was trained using Method 1 (M1) and Method 2 (M2).}
\label{tab:task2}
\small
\begin{tabular}{l|ccc}
\toprule
 & Task 1 (M1) & Task 1 (M2)  & Task 2 \\
 &(EER) & (EER) & (Confusions) \\ \midrule
TNN-cosine & 10.91\% & 8.49\% & \pmb{285} \\
TNN-KNN & 11.72\%. & 8.37\%. &393 \\
TNN-SVM & 11.09\% & 11.35\% &286 \\
baseline  & \pmb{6.24}\% &  N.A  & 369\\\bottomrule
\end{tabular}

\end{table}

While good performance is reached, especially in the case where we are predicting the blacklisted status of known speakers, the power of TNNs lies in classification tasks with few examples per class, as becomes clear when evaluating Task 2.

\subsection{Task 2: Speaker identification}

The limited number of examples available per class (three) makes Task 2 extremely challenging. Our TNN-based methods, however, are able to achieve very high performance. 

When examining the performance on evaluation set A in Table~\ref{tab:task1}, the hybrid TNN-SVM method greatly outperforms all the other methods, resulting in only 358 confusions. In fact, all of the TNN-based methods outperform the baseline. 

Looking at the performance on evaluation set B (Table~\ref{tab:task2}), the method combining TNN with a cosine similarity classifier achieves the best performance with 285 confusions, as compared to a baseline of 369 confusions (23\% improvement). Most of the TNN-based methods are able to outperform the baseline, except for the TNN-KNN. One interesting finding is that, when we train the TNN using training set A (which is a subset of training set B), and evaluate its performance on evaluation set B, we see that it hugely outperforms the baseline results in terms of number of confusions (309 versus 572, 46\% improvement). (Not shown in Table~\ref{tab:task2} to avoid confusion.)



We can thus see that the TNN-based models always outperform the baseline for Task 2, allowing us to conclude that our proposed method is extremely effective for a sparse classification problem with little data per class such as in the case of speaker recognition. In the next subsection, we visualize some of the differences in data distribution between dataset~A and B, so that we can better understand the performance of different models for Task~1.


\subsection{Data Visualization}
\begin{figure}[h]
    \centering
    \includegraphics[width=0.46\textwidth]{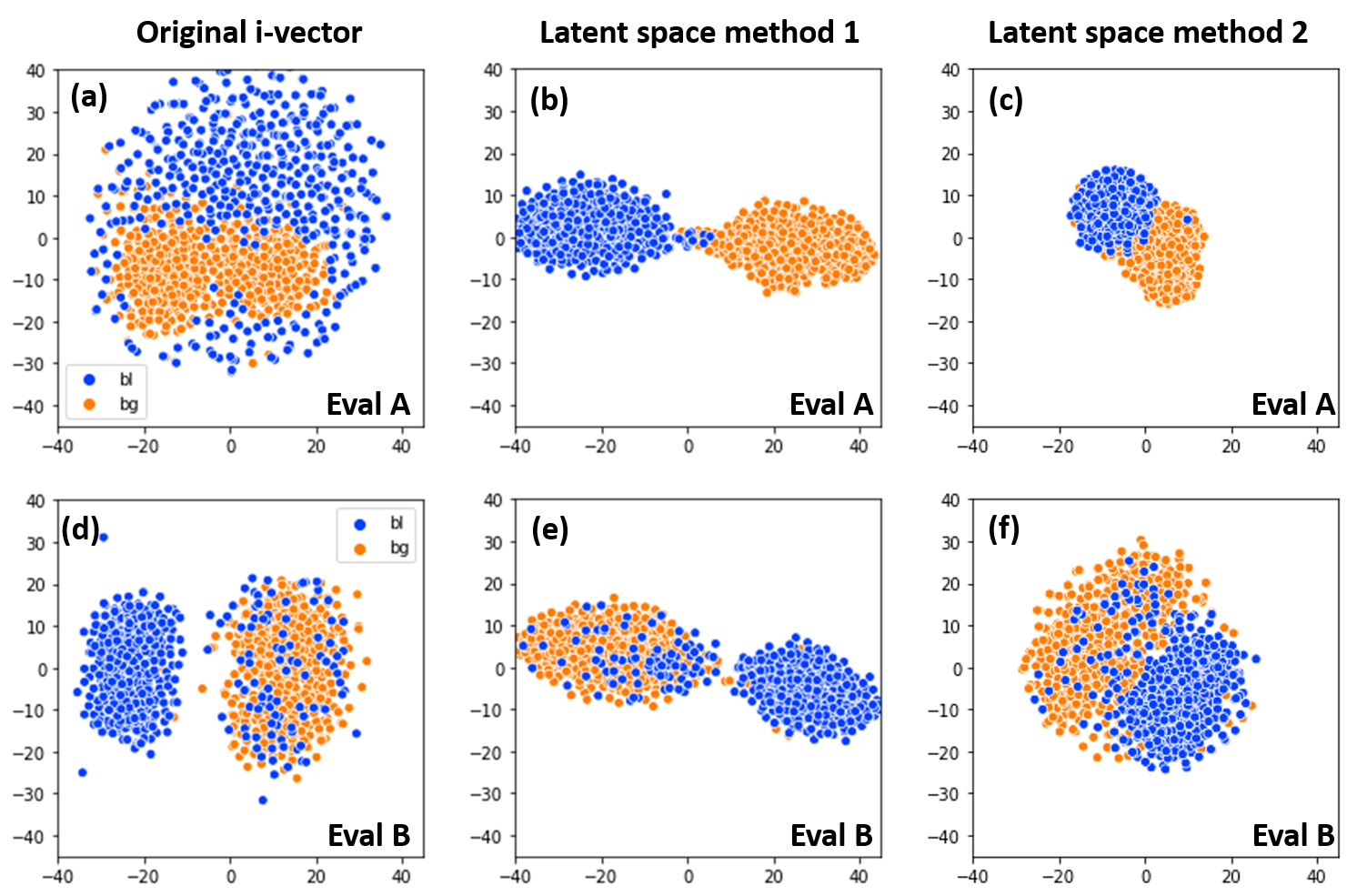}
    \caption{A t-SNE projection of the raw $i$-vectors, and latent space after Method 1 and 2. The colors represent blacklisted and non-blacklisted speakers.}
    \label{fig:task1_visual}
\end{figure}

We visualize the class separation of the raw $i$-vectors and their latent space representations (both Method 1 and Method~2) in Figure~\ref{fig:task1_visual}. A t-Distributed Stochastic Neighbor Embedding (t-SNE) \cite{maaten2008visualizing} was used to project the vector representations in two-dimensional space, with the two colors representing blacklisted versus non-blacklisted speakers (Task 1). We compare the representations for the evaluation set of two datasets (A and B), represented here as evaluation set A (top row) and evaluation set B (bottom row).

When examining Figure~\ref{fig:task1_visual}, we notice that the original $i$-vector representation does not have a clear separation between the blacklisted and non-blacklisted speakers (See Figure~\ref{fig:task1_visual}~(a)). After applying the TNN transformation, the separation between classes becomes much more pronounced (See Figure~\ref{fig:task1_visual} (b) and (c)). 
In particular, after training on training set A, our TNN is able to best separate the two classes, which results in a low EER score for Task 1 (see Table~\ref{tab:task1}). 

When looking at set B, however, we can see that the initial raw data already contains a separate cluster, even though the second cluster still contains mixed data (See Figure~\ref{fig:task1_visual} (d)). When training on this set (B) we get slightly more overlap between the clusters than we did for set A (i.e., comparing  Figure~\ref{fig:task1_visual} (b) and (c) with  Figure~\ref{fig:task1_visual} (e) and (f), respectively), thus visualizing why the EER rate is higher after the TNN transformation for this dataset. The reason for this difference can be found by looking at Table~\ref{tab:datasplit}. In set B, we have the additional challenge that the test set contains unknown non-blacklisted speakers. In future research, we plan on making our hybrid TNN model more robust so that it can better handle unknown speakers. 

When comparing the projections of the original $i$-vectors for set A and B, we see that set A contains more clustered data than set B (See Figure~\ref{fig:task1_visual} (a) and (d)). Despite the cluster to the left for set B, the right half of the figure is very intertwined, thus making it hard to fully separate the classes (See Figure~\ref{fig:task1_visual} (d)). While the TNN performs well for set A, the effect of containing unknown non-blacklisted speakers and the overlapping clusters in set B cause the vectors, even after TNN transformation, to be less separable than the original $i$-vectors for Task 1. This is a topic of further investigation.

\begin{figure}[h]
    \centering
    \includegraphics[width=0.45\textwidth]{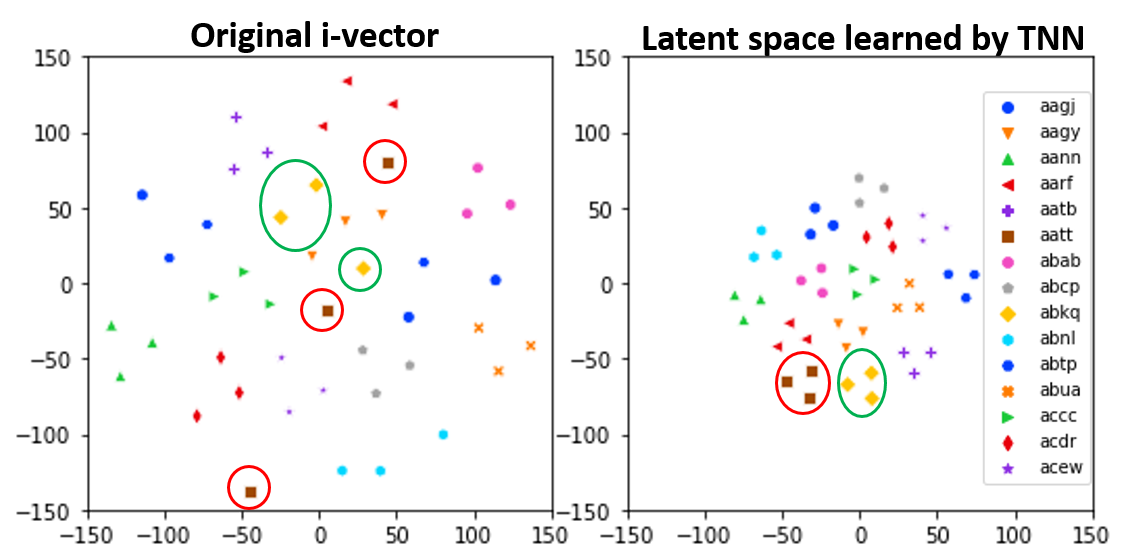}
    \caption{A t-SNE projection of the raw $i$-vectors, and latent space learned by the TNN. The different color/shapes represent different blacklisted speakers. }
    \label{fig:task2_visual}
\end{figure}

The TNN performs very well on Task 2, which is arguably the more difficult task, given that we have only three samples per speaker. Figure~\ref{fig:task2_visual} shows a t-SNE projection of the original $i$-vectors and latent TNN vectors for Task 2. Before the TNN transformation, the original $i$-vector are relatively sparse. Some speakers are even scattered across different classes as indicated by the red and green circles (same color/shape indicates the same speaker). After the TNN transformation, the vectors that belong to the same class are grouped closely together, thus making it easier for a classifier to separate them. 
The resulting hybrid classifier performs very well on the multi-target classification (Task 2), despite the sparsity of the classes.

\subsection{Comparison to other methods}
Our system ranked 7th in speaker recognition and 9th in speaker detection MCE 2018 challenge, out of 12 participants. The top submission obtained 0.86\% EER for Task~1 and 5.96\% for Task~2 (they did not report the number of confusions). Our best performing model, achieves 0.84\% EER for Task~1 and 358 confusions (equivalent to 2.16\% EER) for Task~2. If we only consider training on the training set and validate on development set, our model actually outperforms the other participants. However, when validating on the test set, our model's accuracy is slightly below the best model~\cite{font2019denoising,khourypindrop}. As discussed in Section 4.3, the development and test set distributions are entirely different. Our model has learned to detect and identify speakers given a certain distribution in the training data. Yet, when the test set has a different distribution from the development set, our model does not perform as good, as it violates the basic assumption that the training data and validation data should follow the same statistical distribution. Moreover, the approach used by the top submission is a hybrid system. They use LDA fused with two shallow neural networks. When using only LDA as a stand-alone classifier, it performs good in Task~2 with only 230 confusions (compared to 285 confusions for our model). However, LDA performs poorly in Task~1 with ERR 23.67\% (compared to 8.37\% for our model). Both TNN and LDA seem to work better for different tasks. When LDA is fused with a neural network, it becomes the top submission\cite{font2019denoising,khourypindrop} for MCE 2018. In the future, we may develop a hybrid approach that integrates TNN, as we suspect this may further improve our results just as it did for LDA.

When comparing our system to neural networks with a similar architecture, we achieve better results. For example, \citet{rallabandisubmission}, also used Triplet Neural Networks and Siamese networks. They, however, only obtained 2.35\% EER for Task~1 and 444 confusions for Task~2. Whereas our proposed model reaches a lower EER of 0.84\% on Task~1 and 358 confusions for Task~2, which is 64.3\% (Task~1) and 19.4\% (Task~2) better than Rallabandi et al.'s work. This difference could potentially stem from the fact that they use a more complex neural network architecture (convolutional neural networks), whereas we only use a single layer fully connected network with ReLU activation. This reveals that deeper models do not work well for this (smaller) dataset, potentially due to overfitting.

\section{Conclusion }

We propose a hybrid classification system for speaker detection and identification based on triplet neural networks, which can be downloaded online\footnote{https://github.com/KinWaiCheuk/MCE2018}. The method first learns a new latent space representation of the given $i$-vectors. This new representation is used to train traditional classifiers, and evaluated on the MCE 2018 dataset. 
By transforming the original $i$-vectors into a latent space with a TNN, we achieve a 46\% improvement in classification accuracy relative to the MCE 2018 competition baseline on the speaker identification task (Task 2) when training on a limited dataset (set A). When training on a more extensive dataset (set B, 33\% more data), a 23\% improvement in ERR is achieved, relative to the baseline. This confirms that our approach works well on a small dataset with sparse classes, given that the MCE 2018 dataset contains only three conversations per speaker. 
For Task 1, blacklisted versus non-blacklisted speaker classification, our approach outperforms the baseline when there are no new speakers present in the test set (e.g. evaluation set A). In future research, we plan to improve the ability of the TNN model to generalize for the presence of unknown speakers.  The raw audio files are not given in the dataset, so only $i$-vectors can be studied at the moment. If the raw audio files are available in the future, a comparison of different representation such as $d$-vectors and $x$-vectors can be further investigated.

\section{Acknowledgements}
\label{sec:page}

This work is supported by the SUTD-MIT IDC grant IDG31800103, MOE Grant no. MOE2018-T2-2-161, and SING-2018-02-0204.



\bibliographystyle{IEEEtranN}
\bibliography{strings,refs}

\begin{thebibliography}{33}
\providecommand{\natexlab}[1]{#1}
\providecommand{\url}[1]{#1}
\csname url@samestyle\endcsname
\providecommand{\newblock}{\relax}
\providecommand{\bibinfo}[2]{#2}
\providecommand{\BIBentrySTDinterwordspacing}{\spaceskip=0pt\relax}
\providecommand{\BIBentryALTinterwordstretchfactor}{4}
\providecommand{\BIBentryALTinterwordspacing}{\spaceskip=\fontdimen2\font plus
\BIBentryALTinterwordstretchfactor\fontdimen3\font minus
  \fontdimen4\font\relax}
\providecommand{\BIBforeignlanguage}[2]{{%
\expandafter\ifx\csname l@#1\endcsname\relax
\typeout{** WARNING: IEEEtranN.bst: No hyphenation pattern has been}%
\typeout{** loaded for the language `#1'. Using the pattern for}%
\typeout{** the default language instead.}%
\else
\language=\csname l@#1\endcsname
\fi
#2}}
\providecommand{\BIBdecl}{\relax}
\BIBdecl

\bibitem[Schroff et~al.(2015)Schroff, Kalenichenko, and
  Philbin]{schroff2015facenet}
F.~Schroff, D.~Kalenichenko, and J.~Philbin, ``Facenet: A unified embedding for
  face recognition and clustering,'' in \emph{Proceedings of the IEEE
  conference on computer vision and pattern recognition}, 2015, pp. 815--823.

\bibitem[Ye and Guo(2018)]{ye2018deep}
M.~Ye and Y.~Guo, ``Deep triplet ranking networks for one-shot recognition,''
  \emph{arXiv preprint arXiv:1804.07275}, 2018.

\bibitem[Chen et~al.(2016)Chen, Guo, and Lai]{chen2016deep}
S.-Z. Chen, C.-C. Guo, and J.-H. Lai, ``Deep ranking for person
  re-identification via joint representation learning,'' \emph{IEEE
  Transactions on Image Processing}, vol.~25, no.~5, pp. 2353--2367, 2016.

\bibitem[Cheng et~al.(2016)Cheng, Gong, Zhou, Wang, and Zheng]{cheng2016person}
D.~Cheng, Y.~Gong, S.~Zhou, J.~Wang, and N.~Zheng, ``Person re-identification
  by multi-channel parts-based cnn with improved triplet loss function,'' in
  \emph{Proceedings of the IEEE Conference on Computer Vision and Pattern
  Recognition}, 2016, pp. 1335--1344.

\bibitem[Su et~al.(2016)Su, Zhang, Xing, Gao, and Tian]{su2016deep}
C.~Su, S.~Zhang, J.~Xing, W.~Gao, and Q.~Tian, ``Deep attributes driven
  multi-camera person re-identification,'' in \emph{European conference on
  computer vision}.\hskip 1em plus 0.5em minus 0.4em\relax Springer, 2016, pp.
  475--491.

\bibitem[Wang et~al.(2016)Wang, Zuo, Lin, Zhang, and Zhang]{wang2016joint}
F.~Wang, W.~Zuo, L.~Lin, D.~Zhang, and L.~Zhang, ``Joint learning of
  single-image and cross-image representations for person re-identification,''
  in \emph{Proceedings of the IEEE Conference on Computer Vision and Pattern
  Recognition}, 2016, pp. 1288--1296.

\bibitem[Ding et~al.(2015)Ding, Lin, Wang, and Chao]{ding2015deep}
S.~Ding, L.~Lin, G.~Wang, and H.~Chao, ``Deep feature learning with relative
  distance comparison for person re-identification,'' \emph{Pattern
  Recognition}, vol.~48, no.~10, pp. 2993--3003, 2015.

\bibitem[Song et~al.(2018)Song, Willi, Thiagarajan, Berisha, and
  Spanias]{song2018triplet}
H.~Song, M.~M. Willi, J.~J. Thiagarajan, V.~Berisha, and A.~Spanias, ``Triplet
  network with attention for speaker diarization,'' in \emph{INTERSPEECH},
  2018.

\bibitem[Le~Lan et~al.(2017)Le~Lan, Charlet, Larcher, and
  Meignier]{le2017triplet}
G.~Le~Lan, D.~Charlet, A.~Larcher, and S.~Meignier, ``A triplet ranking-based
  neural network for speaker diarization and linking.'' in \emph{INTERSPEECH},
  2017, pp. 3572--3576.

\bibitem[Zhang et~al.(2018)Zhang, Koishida, and Hansen]{zhang2018text}
C.~Zhang, K.~Koishida, and J.~H. Hansen, ``Text-independent speaker
  verification based on triplet convolutional neural network embeddings,''
  \emph{IEEE/ACM Transactions on Audio, Speech and Language Processing
  (TASLP)}, vol.~26, no.~9, pp. 1633--1644, 2018.

\bibitem[Li et~al.(2018)Li, Tuo, Su, Li, and Yu]{li2018deep}
N.~Li, D.~Tuo, D.~Su, Z.~Li, and D.~Yu, ``Deep discriminative embeddings for
  duration robust speaker verification.'' in \emph{Interspeech}, 2018, pp.
  2262--2266.

\bibitem[Li et~al.(2017)Li, Ma, Jiang, Li, Zhang, Liu, Cao, Kannan, and
  Zhu]{li2017deep}
C.~Li, X.~Ma, B.~Jiang, X.~Li, X.~Zhang, X.~Liu, Y.~Cao, A.~Kannan, and Z.~Zhu,
  ``Deep speaker: an end-to-end neural speaker embedding system,'' \emph{arXiv
  preprint arXiv:1705.02304}, 2017.

\bibitem[Sar{\i} et~al.(2019)Sar{\i}, Thomas, Hasegawa-Johnson, and
  Picheny]{sari2019pre}
L.~Sar{\i}, S.~Thomas, M.~Hasegawa-Johnson, and M.~Picheny, ``Pre-training of
  speaker embeddings for low-latency speaker change detection in broadcast
  news,'' in \emph{ICASSP 2019-2019 IEEE International Conference on Acoustics,
  Speech and Signal Processing (ICASSP)}.\hskip 1em plus 0.5em minus
  0.4em\relax IEEE, 2019, pp. 6286--6290.

\bibitem[Cheuk et~al.(2018)Cheuk, BT, Roig, and Herremans]{cheuk2018mce}
\BIBentryALTinterwordspacing
K.~Cheuk, B.~BT, G.~Roig, and D.~Herremans, ``Blacklisted speaker
  identification using triplet neural networks,'' \emph{{M}{C}{E} 2018: The 1st
  multi-target speaker detection and identification challenge evaluation},
  2018. [Online]. Available:
  \url{http://mce.csail.mit.edu/pdfs/SUTD\_description.pdf}
\BIBentrySTDinterwordspacing

\bibitem[Shon et~al.(2018)Shon, Dehak, Reynolds, and Glass]{shon2018mce}
S.~Shon, N.~Dehak, D.~Reynolds, and J.~Glass, ``Mce 2018: The 1st multi-target
  speaker detection and identification challenge evaluation (mce) plan, dataset
  and baseline system,'' \emph{arXiv preprint arXiv:1807.06663}, 2018.

\bibitem[Hasan and Hansen(2011)]{hasan2011study}
T.~Hasan and J.~H. Hansen, ``A study on universal background model training in
  speaker verification,'' \emph{IEEE Transactions on Audio, Speech, and
  Language Processing}, vol.~19, no.~7, pp. 1890--1899, 2011.

\bibitem[Kenny(2005)]{kenny2005joint}
P.~Kenny, ``Joint factor analysis of speaker and session variability: Theory
  and algorithms,'' \emph{CRIM, Montreal,(Report) CRIM-06/08-13}, vol.~14, pp.
  28--29, 2005.

\bibitem[Soong et~al.(1987)Soong, Rosenberg, Juang, and
  Rabiner]{soong1987report}
F.~K. Soong, A.~E. Rosenberg, B.-H. Juang, and L.~R. Rabiner, ``Report: A
  vector quantization approach to speaker recognition,'' \emph{AT\&T technical
  journal}, vol.~66, no.~2, pp. 14--26, 1987.

\bibitem[Garcia-Romero and Espy-Wilson(2011)]{garcia2011analysis}
D.~Garcia-Romero and C.~Y. Espy-Wilson, ``Analysis of i-vector length
  normalization in speaker recognition systems,'' in \emph{Twelfth Annual
  Conference of the International Speech Communication Association}, 2011.

\bibitem[Izenman(2013)]{izenman2013linear}
A.~J. Izenman, ``Linear discriminant analysis,'' in \emph{Modern multivariate
  statistical techniques}.\hskip 1em plus 0.5em minus 0.4em\relax Springer,
  2013, pp. 237--280.

\bibitem[Ioffe(2006)]{ioffe2006probabilistic}
S.~Ioffe, ``Probabilistic linear discriminant analysis,'' in \emph{European
  Conference on Computer Vision}.\hskip 1em plus 0.5em minus 0.4em\relax
  Springer, 2006, pp. 531--542.

\bibitem[Burget et~al.(2011)Burget, Plchot, Cumani, Glembek, Mat{\v{e}}jka, and
  Br{\"u}mmer]{burget2011discriminatively}
L.~Burget, O.~Plchot, S.~Cumani, O.~Glembek, P.~Mat{\v{e}}jka, and
  N.~Br{\"u}mmer, ``Discriminatively trained probabilistic linear discriminant
  analysis for speaker verification,'' in \emph{Acoustics, Speech and Signal
  Processing (ICASSP), 2011 IEEE International Conference on}.\hskip 1em plus
  0.5em minus 0.4em\relax IEEE, 2011, pp. 4832--4835.

\bibitem[Hoffer and Ailon(2015)]{hoffer2015deep}
E.~Hoffer and N.~Ailon, ``Deep metric learning using triplet network,'' in
  \emph{International Workshop on Similarity-Based Pattern Recognition}.\hskip
  1em plus 0.5em minus 0.4em\relax Springer, 2015, pp. 84--92.

\bibitem[Cheuk et~al.()Cheuk, Balamurali, Roig, and
  Herremans]{cheukblacklisted}
K.~W. Cheuk, B.~Balamurali, G.~Roig, and D.~Herremans, ``Blacklisted speaker
  identification using triplet neural networks,'' \emph{MCE 2018 Challenge
  System Description}, available at
  \url{http://mce.csail.mit.edu/pdfs/SUTD\_description.pdf}.

\bibitem[Hearst et~al.(1998)Hearst, Dumais, Osuna, Platt, and
  Scholkopf]{hearst1998support}
M.~A. Hearst, S.~T. Dumais, E.~Osuna, J.~Platt, and B.~Scholkopf, ``Support
  vector machines,'' \emph{IEEE Intelligent Systems and their applications},
  vol.~13, no.~4, pp. 18--28, 1998.

\bibitem[Steinwart and Christmann(2008)]{steinwart2008support}
I.~Steinwart and A.~Christmann, \emph{Support vector machines}.\hskip 1em plus
  0.5em minus 0.4em\relax Springer Science \& Business Media, 2008.

\bibitem[Keller et~al.(1985)Keller, Gray, and Givens]{keller1985fuzzy}
J.~M. Keller, M.~R. Gray, and J.~A. Givens, ``A fuzzy k-nearest neighbor
  algorithm,'' \emph{IEEE transactions on systems, man, and cybernetics},
  no.~4, pp. 580--585, 1985.

\bibitem[Weinberger et~al.(2006)Weinberger, Blitzer, and
  Saul]{weinberger2006distance}
K.~Q. Weinberger, J.~Blitzer, and L.~K. Saul, ``Distance metric learning for
  large margin nearest neighbor classification,'' in \emph{Advances in neural
  information processing systems}, 2006, pp. 1473--1480.

\bibitem[Pedregosa et~al.(2011)Pedregosa, Varoquaux, Gramfort, Michel, Thirion,
  Grisel, Blondel, Prettenhofer, Weiss, Dubourg, Vanderplas, Passos,
  Cournapeau, Brucher, Perrot, and Duchesnay]{scikit-learn}
F.~Pedregosa, G.~Varoquaux, A.~Gramfort, V.~Michel, B.~Thirion, O.~Grisel,
  M.~Blondel, P.~Prettenhofer, R.~Weiss, V.~Dubourg, J.~Vanderplas, A.~Passos,
  D.~Cournapeau, M.~Brucher, M.~Perrot, and E.~Duchesnay, ``Scikit-learn:
  Machine learning in {P}ython,'' \emph{Journal of Machine Learning Research},
  vol.~12, pp. 2825--2830, 2011.

\bibitem[Maaten and Hinton(2008)]{maaten2008visualizing}
L.~v.~d. Maaten and G.~Hinton, ``Visualizing data using t-sne,'' \emph{Journal
  of machine learning research}, vol.~9, no. Nov, pp. 2579--2605, 2008.

\bibitem[Font(2019)]{font2019denoising}
R.~Font, ``A denoising autoencoder for speaker recognition. results on the mce
  2018 challenge,'' in \emph{ICASSP 2019-2019 IEEE International Conference on
  Acoustics, Speech and Signal Processing (ICASSP)}.\hskip 1em plus 0.5em minus
  0.4em\relax IEEE, 2019, pp. 6016--6020.

\bibitem[Khoury et~al.()Khoury, Lakhdhar, Vaughan, Sivaraman, and
  Nagarsheth]{khourypindrop}
E.~Khoury, K.~Lakhdhar, A.~Vaughan, G.~Sivaraman, and P.~Nagarsheth, ``Pindrop
  submission to mce 2018.''

\bibitem[Rallabandi and Black()]{rallabandisubmission}
S.~Rallabandi and A.~W. Black, ``Submission from cmu towards 1st multitarget
  speaker detection and identification challenge.''

\end{thebibliography}

\end{document}